\documentstyle[preprint,pra,aps]{revtex}

\begin{document}

\draft

\title{Theory of Interacting Quantum Gases}

\author{H.T.C. Stoof, M. Bijlsma, and M. Houbiers}
\address{University of Utrecht, Institute for Theoretical
         Physics, Princetonplein 5, \\
         P.O. Box 80.006, 3508 TA  Utrecht, The Netherlands}

\maketitle

\begin{abstract}
We present a unified picture of the interaction effects in dilute atomic
quantum gases. We consider fermionic as well as bosonic gases and, in
particular, discuss for both forms of statistics the fundamental differences
between a gas with effectively repulsive and a gas with effectively attractive
interatomic interactions, i.e.\ between a gas with either a positive or a
negative scattering length.
\end{abstract}

\pacs{\\ PACS numbers: 03.75.Fi, 32.80.Pj, 42.50.Vk, 67.65.+z}

\section{INTRODUCTION}
\label{int}
One of the most important objectives of atomic physics in the last two decades,
has been the achievement of Bose-Einstein condensation in a weakly interacting
gas. It is, therefore, not surprising that the observation of Bose-Einstein
condensation in the magnetically trapped alkali vapors $^{87}$Rb \cite{JILA},
$^7$Li \cite{rice}, and $^{23}$Na \cite{MIT}, is considered to be one of the
most exciting events of the last year. This is not in the least the case,
because now that the successful road towards Bose-Einstein condensation has
been found, one can safely assume that gases of atomic $^{85}$Rb, $^{133}$Cs,
and $^1$H will also be Bose condensed in the near future and that a large
number of different experimental systems will be available to study various
aspects of the condensation phenomenon in detail.

Moreover, both lithium and hydrogen have stable fermionic isotopes which can be
trapped and cooled in a similar manner as their bosonic counterparts. For
atomic $^6$Li this has already been achieved \cite{randy}, but magnetically
trapped deuterium ($^2$H) has not been observed yet because the loading of the
trap cannot be accomplished in the same way as in the case of atomic hydrogen.
This is presumably caused by the fact that i) deuterium binds more strongly to
a superfluid helium film, ii) the surface recombination rate is much larger,
and iii) the sample is contaminated with atomic hydrogen \cite{meritt}.
Nevertheless, it is to be expected that these experimental difficulties can be
overcome and that one will soon be able to study both these fermion gases in
the degenerate regime.

In view of the exciting developments mentioned above, it appears justified to
present here an overview of the physical properties of all these weakly
interacting quantum gases. We thereby intend to bring out most clearly the
differences due to the Fermi or Bose statistics on the one hand, and due to the
effective interatomic interaction being either repulsive or attractive on the
other. This division in the focus of the paper is also reflected in its
lay-out. In Sec.~\ref{quantum} we present a general introduction into the
physics of the atomic quantum gases of interest. In particular, we specify the
approximations that are allowed for these quantum systems, and subsequently
show how one can determine the effective interaction between the atoms. Using
the latter result, we consider in Sec.~\ref{fermi} the Fermi and in
Sec.~\ref{bose} the Bose gases. Because we wish to bring out the different
physical properties most clearly, we treat there only the homogeneous case. The
experimentally more realistic case of a gas in a harmonic oscillator potential,
however, does not lead to any qualitative differences, except for a Bose gas
with an effectively attractive interaction. The latter possibility is therefore
briefly discussed in Sec.~\ref{out}, where we also end the paper by pointing
out some problems that remain for the future.

\section{INTERACTING QUANTUM GASES}
\label{quantum}
All atoms mentioned in the introduction are members of the first primary group
of the Periodic Table of Elements and, therefore, have a magnetic moment
$\vec{\mu}$ that is equal to the sum of the magnetic moments of the electron
spin $\vec{s}$ (with $s=1/2$) and the nuclear spin $\vec{i}$. Due to this
magnetic moment, the atoms can easily be trapped by an inhomogeneous magnetic
field and can be evaporatively cooled by a resonant microwave field that flips
the magnetic moment. The successful experiments of last year have shown that
this is a great advantage. However, the magnetic moment also gives rise to an
important problem because two-body collisions in the gas can lead to a change
in the component of the magnetic moment parallel to the magnetic field and thus
to a decay of the atomic density in the magnetic trap.

This decay mechanism seriously limits the lifetime of the gas, and can in
principle be caused both directly by the magnetic dipole-dipole interaction
$V^d$ and by the central (singlet/triplet) interaction $V^c$ in combination
with the hyperfine interaction $V^{hf} = a_{hf} \vec{s} \cdot \vec{i}$
\cite{H}. In general, the typical timescale for the former process is much
larger than that for the latter, and the gas spontaneously spin-polarizes to a
state in which the lifetime $\tau_{inel}$ is determined solely by the magnetic
dipole-dipole interaction. There are roughly speaking two scenarios: At small
magnetic fields $B \ll a_{hf}/\mu_e$, where $\mu_e$ denotes the electron
magneton, the decay due to the central interaction can only be avoided by
polarizing both the electron and the nuclear spin along the direction of the
magnetic field, i.e.\ all the atoms are in the spin state $|m_s=1/2,
m_i=i\rangle$. This corresponds to a so-called doubly spin-polarized atomic
gas. However, for large magnetic fields $B \gg a_{hf}/\mu_e$, the nuclear spin
does not necessarily have to be polarized to obtain a relatively long lifetime
and one can essentially trap any mixture of the states $|m_s=1/2, m_i\rangle$
with an arbitrary projection of the nuclear spin. In this case we are dealing
with an (electron) spin-polarized atomic gas.

It is important to note that although the central interaction does not
determine the lifetime of the (doubly) spin-polarized gas, it does determine
the typical timescale $\tau_{el}$ for elastic collisions. (See, however,
Sec.~\ref{fermi} for one exception.) Because $\tau_{el} \ll \tau_{inel}$ for
not too low temperatures, we can consider the spatial degrees of freedom of the
gas to be in equilibrium, whereas the spin degrees of freedom clearly are not.
For timescales short compared to the lifetime, we are thus allowed to apply
equilibrium statistical mechanics to the gas and our main problem is the
accurate treatment of the elastic part of the central interaction, which for
the (doubly) spin-polarized gases of interest is the triplet interaction $V_T$.

The determination of the effect of the interaction on the thermodynamic
properties of a quantum gas is considerably simplified by the presence of two
small parameters in the problem. In terms of density $n$ of the gas, the range
of the interaction $r_V$, and the thermal de Broglie wavelength
$\Lambda = \sqrt{2\pi\hbar^2/mk_BT}$ of an atom with mass $m$, the two small
parameters are the gas parameter $nr_V^3$ and the quantum parameter
$r_V/\Lambda$. Physically, the condition $nr_V^3 \ll 1$ expresses the fact that
we are dealing with a dilute system in which it is highly improbable that three
particles are so close together that they can interact with each other
simultaneously. As a result we can neglect three-body processes and only have
to account for all two-body processes. Furthermore, the condition $r_V/\Lambda
\ll 1$ shows that if two atoms interact, their relative angular momentum can
only be equal to zero. In combination, we therefore conclude that for an
accurate description of a quantum gas we only need to consider all possible
two-body s-wave scattering processes.

\subsection{Two-body transition matrix}
\label{two}
To explain how this may be achieved, we first of all introduce the appropriate
`grand-canonical' hamiltonian of the (doubly) spin-polarized atomic gas. Using
the language of second quantization \cite{fetter}, it reads
\begin{equation}
\label{hamilton}
H = \sum_{\alpha} \sum_{\vec{k}}~
     (\epsilon_{\vec{k},\alpha} - \mu_{\alpha})
        a^{\dagger}_{\vec{k},\alpha} a_{\vec{k},\alpha}
      + \frac{1}{2V} \sum_{\alpha \alpha'}
                         \sum_{\vec{k} \vec{k}' \vec{q}}~
           V_{\vec{q}}~a^{\dagger}_{\vec{k}+\vec{q},\alpha}
                       a^{\dagger}_{\vec{k}'-\vec{q},\alpha'}
                       a_{\vec{k}',\alpha'} a_{\vec{k},\alpha}
  \equiv H_0 + H_{int}~,
\end{equation}
where $a^{\dagger}_{\vec{k},\alpha}$ creates and $a_{\vec{k},\alpha}$
annihilates an atom in the (electron and nuclear) spin state $|\alpha\rangle$
and the momentum eigenstate $|\vec{k}\rangle$ of a cubic box with volume $V$
and periodic boundary conditions. As always, these creation and annihilation
operators obey the (anti)commutation relations
$[a_{\vec{k},\alpha},a^{\dagger}_{\vec{k}',\alpha'}]_{\pm} =
              \delta_{\vec{k},\vec{k}'} \delta_{\alpha,\alpha'}$
and
$[a_{\vec{k},\alpha},a_{\vec{k}',\alpha'}]_{\pm} =
[a^{\dagger}_{\vec{k},\alpha},
                     a^{\dagger}_{\vec{k}',\alpha'}]_{\pm} = 0$ if the atoms
are fermions or bosons, respectively.
Moreover, $V_{\vec{q}} = \int d\vec{x}~V_T(\vec{x})
                                  e^{-i \vec{q} \cdot \vec{x}}$
denotes the Fourier transform of the triplet interaction and
$\epsilon_{\vec{k},\alpha} =
         \hbar^2 \vec{k}^2/2m + \epsilon_{\alpha}
         \equiv \epsilon_{\vec{k}} + \epsilon_{\alpha} $
is the energy of the single particle states $|\vec{k},\alpha\rangle$. Note that
both the spin state $|\alpha\rangle$ and its energy $\epsilon_{\alpha}$ depend
on the applied magnetic field $\vec{B}$ due to the Zeeman interaction
$V^z = - \vec{\mu} \cdot \vec{B}$. This dependence can for any particular atom
easily be determined by a diagonalization of the spin part of the atomic
hamiltonian, i.e\ of $V^{hf} + V^z$, but is not explicitly calculated here
since it plays only a minor role in the following. Note also that in agreement
with our previous remarks, we have only assumed that the spatial degrees of
freedom are in equilibrium and have, therefore, introduced a different chemical
potential $\mu_{\alpha}$ for each spin state $|\alpha\rangle$.

Next we consider the effect of the interaction $H_{int}$ on an `initial' state
$|i\rangle$ of two atoms. In particular, we take $|i\rangle$ to be the state in
which the occupation numbers of all the single particle states are zero except
for
$N_{\vec{K}/2 + \vec{k},\alpha}$ and
$N_{\vec{K}/2 - \vec{k},\alpha'}$, which are equal to one. In lowest order in
the interaction, the energy of this state is
$E_i =\epsilon_{\vec{K}/2 + \vec{k},\alpha} +
                       \epsilon_{\vec{K}/2 - \vec{k},\alpha'}$
if we use that the chemical potentials for this two-body problem are equal to
zero. Applying standard perturbation theory, we find that this energy is
shifted due to the interaction by an amount
\begin{equation}
\Delta E_i = \langle i|H_{int}|i \rangle +
   \sum_{f \neq i}~ \langle i|H_{int}|f \rangle
                      \frac{1}{E_i - E_f}
                   \langle f|H_{int}|i \rangle + \dots~,
\end{equation}
where the `final' state $|f\rangle$ is an eigenstate of $H_0$ with energy
$E_f$. Moreover, due to the possibility of scattering to other momentum states,
the state $|i\rangle$ also acquires a finite lifetime $\tau_i$ that up to
second order perturbation theory is found from Fermi's Golden Rule
\begin{equation}
\frac{1}{\tau_i} = \frac{2\pi}{\hbar}
   \sum_{f}~ \delta(E_i - E_f)
                    |\langle f|H_{int}|i \rangle|^2~.
\end{equation}
Introducing the usual notation $1/(x + i0)$ for the limiting procedure
$\lim_{\eta \downarrow 0}~1/(x + i\eta)$, both these results can be
conveniently combined into
\begin{equation}
\label{pert}
\Delta E_i - \frac{i\hbar}{2\tau_i} =
   \langle i|H_{int}|i \rangle +
   \sum_{f}~ \langle i|H_{int}|f \rangle
                      \frac{1}{E_i - E_f + i0}
                   \langle f|H_{int}|i \rangle + \dots~.
\end{equation}
The relevant matrixelements of the interaction $H_{int}$ are easily evaluated
and we obtain that
\begin{eqnarray}
\Delta E_i - \frac{i\hbar}{2\tau_i} =
  \frac{1}{V} \left\{ V_{\vec{0}} +
     \frac{1}{V} \sum_{\vec{k}'}~
             V_{\vec{k}-\vec{k}'}
               \frac{1}{\hbar^2(\vec{k}^2 - \vec{k}'^2)/m + i0}
             V_{\vec{k}'-\vec{k}} + \dots \right. \hspace*{1.0in}
                                                     \nonumber \\
    \mp \left. \delta_{\alpha,\alpha'}
              \left( V_{-2\vec{k}} +
     \frac{1}{V} \sum_{\vec{k}'}~
             V_{-\vec{k}-\vec{k}'}
               \frac{1}{\hbar^2(\vec{k}^2 - \vec{k}'^2)/m + i0}
             V_{\vec{k}'-\vec{k}} + \dots \right)
              \right\}~,
\end{eqnarray}
which in the continuum limit $V \rightarrow \infty$ becomes
\begin{eqnarray}
\label{energy}
\Delta E_i - \frac{i\hbar}{2\tau_i} =
  \frac{1}{V} \left\{ V(\vec{0}) +
     \int \frac{d\vec{k}'}{(2\pi)^3}~
             V(\vec{k}-\vec{k}')
               \frac{1}{\hbar^2(\vec{k}^2 - \vec{k}'^2)/m + i0}
             V(\vec{k}'-\vec{k}) + \dots \right. \hspace*{0.3in}
                                                     \nonumber \\
    \mp \left. \delta_{\alpha,\alpha'}
              \left( V(-2\vec{k}) +
        \int \frac{d\vec{k}'}{(2\pi)^3}~
             V(-\vec{k}-\vec{k}')
               \frac{1}{\hbar^2(\vec{k}^2 - \vec{k}'^2)/m + i0}
             V(\vec{k}'-\vec{k}) + \dots \right)
              \right\}~.
\end{eqnarray}

The higher order contributions can be calculated in a similar way and we
actually find that the complete perturbation series can be summed by
introducing the so-called two-body T(ransition) matrix
$T^{2B}(\vec{k}',\vec{k};E)$, which obeys the famous Lippman-Schwinger equation
\cite{walter}
\begin{equation}
\label{t2b}
T^{2B}(\vec{k}',\vec{k};E) = V(\vec{k}'-\vec{k}) +
     \int \frac{d\vec{k}''}{(2\pi)^3}~
             V(\vec{k}'-\vec{k}'')
               \frac{1}{E - \hbar^2\vec{k}''^2/m + i0}
                                    T^{2B}(\vec{k}'',\vec{k};E)
\end{equation}
and is proportional to the amplitude for two atoms to collide with a kinetic
energy $E$ in the center-of-mass frame and change their relative momentum from
$\hbar\vec{k}$ to $\hbar\vec{k}'$. Indeed, a comparison with Eq.~(\ref{energy})
shows that the exact result is
\begin{equation}
\Delta E_i - \frac{i\hbar}{2\tau_i} =
  \frac{1}{V} \left\{
             T^{2B}(\vec{k},\vec{k};\hbar^2\vec{k}^2/m) \mp
    \delta_{\alpha,\alpha'}
             T^{2B}(-\vec{k},\vec{k};\hbar^2\vec{k}^2/m)
                     \right\}~.
\end{equation}
We therefore conclude that for a system of two atoms we can account for all
two-body processes by first replacing in the hamiltonian of
Eq.~(\ref{hamilton}) the potential $V_{\vec{q}}$ by the transition matrix
$T^{2B}((\vec{k}-\vec{k}')/2 + \vec{q}, (\vec{k}-\vec{k}')/2;
                              \hbar^2(\vec{k}-\vec{k}')^2/4m)$
and then using only first-order perturbation theory to avoid a double counting
of the effects of the interaction.

\subsection{Many-body transition matrix}
\label{many}
In a gas we are of course not dealing with only two atoms, but with a large
number $N=nV$ of atoms at a temperature $T=1/k_B\beta$. This implies that we
should repeat our previous calculation for a grand canonical ensemble of
`initial' states, in which the single particle states $|\vec{k},\alpha\rangle$
have the average occupation numbers
$N_{\vec{k},\alpha} =
 (\exp\beta(\epsilon_{\vec{k},\alpha}-\mu_{\alpha}) \pm 1)^{-1}$ and the
chemical potentials $\mu_{\alpha}$ are determined from
\begin{equation}
N = \sum_{\alpha} \sum_{\vec{k}}~ N_{\vec{k},\alpha}
  \equiv \sum_{\alpha}~N_{\alpha}~.
\end{equation}
The average zeroth-order energy in this ensemble is equal to
\begin{equation}
E_i = \sum_{\alpha} \sum_{\vec{k}}~
        (\epsilon_{\vec{k},\alpha} - \mu_{\alpha})
                                          N_{\vec{k},\alpha}
\end{equation}
and its average energy shift and lifetime due to the interaction $H_{int}$ can
now be calculated perturbatively by averaging Eq.~(\ref{pert}). Performing this
thermal average is somewhat cumbersome, but ultimately we find
\begin{eqnarray}
\Delta E_i &-& \frac{i\hbar}{2\tau_i} =
  \frac{1}{2V} \sum_{\alpha \alpha'} \sum_{\vec{K} \vec{k}}~
   N_{\vec{K}/2 + \vec{k},\alpha}
   N_{\vec{K}/2 - \vec{k},\alpha'}                 \nonumber \\
    &\times& \left\{ V_{\vec{0}} + \frac{1}{V} \sum_{\vec{k}'}~
             V_{\vec{k}-\vec{k}'}
               \frac{1 \mp N_{\vec{K}/2+\vec{k}',\alpha}
                       \mp N_{\vec{K}/2-\vec{k}',\alpha'}}
                    {\hbar^2(\vec{k}^2 - \vec{k}'^2)/m + i0}
             V_{\vec{k}'-\vec{k}} + \dots \right.  \nonumber \\
& &\hspace*{0.4in} \mp \left. \delta_{\alpha,\alpha'}
              \left( V_{-2\vec{k}} +
        \frac{1}{V} \sum_{\vec{k}'}~
             V_{-\vec{k}-\vec{k}'}
               \frac{1 \mp N_{\vec{K}/2+\vec{k}',\alpha}
                       \mp N_{\vec{K}/2-\vec{k}',\alpha'}}
                    {\hbar^2(\vec{k}^2 - \vec{k}'^2)/m + i0}
             V_{\vec{k}'-\vec{k}} + \dots \right)
              \right\}~,
\end{eqnarray}
neglecting between the curly brackets terms that are either of order
$O(V_{\vec{0}}~(nr_V^3)^{1/3})$ or of order $O(V_{\vec{0}}~r_V/\Lambda)$.
(Diagrammatically speaking, we neglect the four `bubble' diagrams compared to
the two `ladder' diagrams.)

The appearance of the factor
$1 \mp N_{\vec{K}/2+\vec{k}',\alpha}
                            \mp N_{\vec{K}/2-\vec{k}',\alpha'}$ in this
expression may appear surprising at first instance. It has, however, a clear
physical meaning that becomes evident when we consider the contribution to the
average lifetime $\tau_i$ from a scattering process with $\alpha \neq \alpha'$.
Such a contribution is given by
\begin{eqnarray}
\frac{2\pi}{\hbar} \frac{1}{V} \sum_{\vec{k}'}~
  \delta\left( \frac{\hbar^2\vec{k}^2}{m} -
                          \frac{\hbar^2\vec{k}'^2}{m} \right)
     |V_{\vec{k}-\vec{k}'}|^2
       N_{\vec{K}/2 + \vec{k},\alpha}
       N_{\vec{K}/2 - \vec{k},\alpha'}
               \left(1 \mp N_{\vec{K}/2+\vec{k}',\alpha}
                       \mp N_{\vec{K}/2-\vec{k}',\alpha'} \right)
                                                       \nonumber
\end{eqnarray}
and consists of the sum of a loss rate
\begin{eqnarray}
\frac{2\pi}{\hbar} \frac{1}{V} \sum_{\vec{k}'}~
  \delta\left( \frac{\hbar^2\vec{k}^2}{m} -
                          \frac{\hbar^2\vec{k}'^2}{m} \right)
     |V_{\vec{k}-\vec{k}'}|^2
     N_{\vec{K}/2 + \vec{k},\alpha}
     N_{\vec{K}/2 - \vec{k},\alpha'}
             \left( 1 \mp N_{\vec{K}/2+\vec{k}',\alpha} \right)
             \left( 1 \mp N_{\vec{K}/2-\vec{k}',\alpha'} \right)
                                                       \nonumber
\end{eqnarray}
due to the collision of two particles, and a production rate
\begin{eqnarray}
- \frac{2\pi}{\hbar} \frac{1}{V} \sum_{\vec{k}'}~
    \delta\left( \frac{\hbar^2\vec{k}^2}{m} -
                          \frac{\hbar^2\vec{k}'^2}{m} \right)
     |V_{\vec{k}-\vec{k}'}|^2
       N_{\vec{K}/2 + \vec{k},\alpha}
       N_{\vec{K}/2 - \vec{k},\alpha'}
               N_{\vec{K}/2+\vec{k}',\alpha}
               N_{\vec{K}/2-\vec{k}',\alpha'}          \nonumber
\end{eqnarray}
due to the collision of two holes. For a scattering process with $\alpha =
\alpha'$ we can obtain a similar interpretation, although it is slightly more
complicated in this case due to the interference effects.

Following our discussion of the two-atom problem in Sec.~\ref{two}, we see that
we can now include all two-body processes by introducing a many-body
T(ransition) matrix $T^{MB}_{\alpha,\alpha'}(\vec{k}',\vec{k},\vec{K};E)$ that
obeys
\begin{eqnarray}
T^{MB}_{\alpha,\alpha'}(\vec{k}',\vec{k},\vec{K};E) =
  V(\vec{k}'-\vec{k})              \hspace*{3.5in} \nonumber \\
   + \int \frac{d\vec{k}''}{(2\pi)^3}~
           V(\vec{k}'-\vec{k}'')
               \frac{1 \mp N_{\alpha}(\vec{K}/2+\vec{k}'')
                       \mp N_{\alpha'}(\vec{K}/2-\vec{k}'')}
                    {E - \hbar^2\vec{k}''^2/m + i0}
           T^{MB}_{\alpha,\alpha'}(\vec{k}'',\vec{k},\vec{K};E)
\end{eqnarray}
in the continuum limit. Using Eq.~(\ref{t2b}), this is equivalent to
\begin{eqnarray}
\label{tmb}
T^{MB}_{\alpha,\alpha'}(\vec{k}',\vec{k},\vec{K};E) =
  T^{2B}(\vec{k}',\vec{k};E)   \hspace*{3.3in} \nonumber \\
   \mp \int \frac{d\vec{k}''}{(2\pi)^3}~
           T^{2B}(\vec{k}',\vec{k}'';E)
               \frac{N_{\alpha}(\vec{K}/2+\vec{k}'')
                     + N_{\alpha'}(\vec{K}/2-\vec{k}'')}
                    {E - \hbar^2\vec{k}''^2/m + i0}
           T^{MB}_{\alpha,\alpha'}(\vec{k}'',\vec{k},\vec{K};E)~,
\end{eqnarray}
which is more convenient for our purposes since it offers the opportunity to
explore the consequences of the small parameters $nr_V^3$ and $r_V/\Lambda$.
Indeed, as a result of these small parameters the relevant momenta and energies
are always small compared to $\hbar/r_V$ and $\hbar^2/mr_V^2$, respectively,
and we can safely neglect the momentum and energy dependence of the two-body T
matrix. Consequently, Eq.~(\ref{tmb}) is immediately solved by
\begin{equation}
\label{Txi}
T^{MB}_{\alpha,\alpha'}(\vec{k}',\vec{k},\vec{K};E) =
\frac{T^{2B}(\vec{0},\vec{0};0)}
     {1 \pm T^{2B}(\vec{0},\vec{0};0)
                            \Xi_{\alpha,\alpha'}(\vec{K};E)}~,
\end{equation}
where the quantity $T^{2B}(\vec{0},\vec{0};0) = 4\pi a\hbar^2/m$ is
proportional to the (triplet) s-wave scattering length $a$ and
$\Xi_{\alpha,\alpha'}(\vec{K};E)$ is defined by
\begin{equation}
\label{xi}
\Xi_{\alpha,\alpha'}(\vec{K};E) \equiv
   \int \frac{d\vec{k}''}{(2\pi)^3}~
               \frac{N_{\alpha}(\vec{K}/2+\vec{k}'')
                     + N_{\alpha'}(\vec{K}/2-\vec{k}'')}
                    {E - \hbar^2\vec{k}''^2/m + i0}~.
\end{equation}
Moreover, using this solution we find that in a good approximation
\begin{eqnarray}
\label{de}
\Delta E_i - \frac{i\hbar}{2\tau_i} =
  \frac{1}{2V} \sum_{\alpha \alpha'} \sum_{\vec{K} \vec{k}}~
     N_{\vec{K}/2 + \vec{k},\alpha}
     N_{\vec{K}/2 - \vec{k},\alpha'}
        (1 \mp \delta_{\alpha,\alpha'})
        T^{MB}_{\alpha,\alpha'}(\vec{0},\vec{0},\vec{K};
           \hbar^2\vec{k}^2/m)~,
\end{eqnarray}
which can again be reproduced by first-order perturbation theory if we replace
$V_{\vec{q}}$ in Eq.~(\ref{hamilton}) by
$T^{MB}_{\alpha,\alpha'}(\vec{0},\vec{0},\vec{k}+\vec{k}';
 \hbar^2(\vec{k}-\vec{k}')^2/4m)$.

With the last conclusion we have basically achieved our objective of summing
all two-body s-wave scattering processes. In summary, we have shown that this
requires the introduction of a many-body T matrix, which physically also
incorporates the fact that the collisions of interest take place in a (gaseous)
medium and are, therefore, influenced by the usual Fermi-blocking or
Bose-enhancement factors \cite{fetter}. As a result, the transition matrix
$T^{MB}_{\alpha,\alpha'}(\vec{k}',\vec{k},\vec{K};E)$ acquires a dependence on
the center-of-mass momentum $\hbar\vec{K}$ and the energy $E$ that is
explicitly displayed in Eqs.~(\ref{Txi}) and (\ref{xi}). This turns out to be
important for the rest of the paper, where we use the many-body T matrix to
arrive at an accurate (selfconsistent) theory of interacting quantum gases.

\section{FERMI GASES}
\label{fermi}
We start our discussion of the interacting quantum gases with the Fermi case.
We will only consider spin-polarized atomic gases in which two hyperfine states
are almost equally populated. For convenience, these two hyperfine states will
be denoted by $|\uparrow\rangle$ and $|\downarrow\rangle$, respectively. The
reason why we do not discuss doubly spin-polarized gases, is that in such gases
the atoms can only scatter via p-waves due to the Pauli principle. As a result
the (thermalizing) elastic collisions are dominated by the long-range magnetic
dipole-dipole interaction $V^d$ and the spatial degrees of freedom only
equilibrate on a timescale that is comparable to the lifetime of the gas, i.e.\
$\tau_{el} \simeq \tau_{inel}$. Furthermore, if one is able to cool the system
by other means, the gas basically behaves as an ideal Fermi gas on timescales
short compared to the lifetime and interaction effects are negligible (cf.\
Eq.~(\ref{de})).

\subsection{The case ${\bf a>0}$}
\label{Fg0}
We have argued that to account for all two-body s-wave scattering processes we
must use the interaction
\begin{equation}
H_{int} = \frac{1}{V} \sum_{\vec{k} \vec{k}' \vec{q}}~
      T^{MB}_{\uparrow,\downarrow}
        (\vec{0},\vec{0},\vec{k}+\vec{k}';
         \hbar^2(\vec{k}-\vec{k}')^2/4m)~
      a^{\dagger}_{\vec{k}+\vec{q},\uparrow}
      a^{\dagger}_{\vec{k}'-\vec{q},\downarrow}
      a_{\vec{k}',\downarrow} a_{\vec{k},\uparrow}~.
\end{equation}
However, this is not completely true because this interaction induces on
average also a change in the hamiltonian $H_0$, due to the fact that the
occupation numbers
$N_{\vec{k},\alpha} = \langle a^{\dagger}_{\vec{k},\alpha}
                              a_{\vec{k},\alpha} \rangle$ do not vanish. This
so-called mean-field correction is clearly given by
\begin{eqnarray}
\Delta H_0 = \sum_{\alpha} \sum_{\vec{k}}~
    a^{\dagger}_{\vec{k},\alpha} a_{\vec{k},\alpha}
      \left\{ \frac{1}{V} \sum_{\alpha' \neq \alpha}
                                          \sum_{\vec{k}'}~
        T^{MB}_{\uparrow,\downarrow}
         (\vec{0},\vec{0},\vec{k}+\vec{k}';
          \hbar^2(\vec{k}-\vec{k}')^2/4m)
        N_{\vec{k}',\alpha'} \right\}
\end{eqnarray}
and, denoting the density of atoms in the spin state $|\alpha\rangle$ by
$n_{\alpha}$, is well approximated by
\begin{equation}
\label{shift}
\Delta H_0 = \sum_{\alpha} \sum_{\vec{k}}~
    a^{\dagger}_{\vec{k},\alpha} a_{\vec{k},\alpha}
      \left\{ \sum_{\alpha' \neq \alpha}~
        T^{2B}(\vec{0},\vec{0};0) n_{\alpha'} \right\}~,
\end{equation}
since the dominant contributions to the integration over the momentum
$\hbar\vec{k}'$ come from the region near the Fermi momentum
$\hbar k_{F,\alpha'} \equiv
                     \sqrt{2m(\mu_{\alpha'}-\epsilon_{\alpha'})}$
(or from the thermal momentum $\hbar/\Lambda$ if the gas is not degenerate)
where the many-body T matrix is almost equal to the two-body T matrix. The
change in $H_0$ can, therefore, be absorbed in a suitable redefinition of the
chemical potentials. Introducing
\begin{equation}
\mu'_{\alpha} = \mu_{\alpha} - \sum_{\alpha' \neq \alpha}
        T^{2B}(\vec{0},\vec{0};0) n_{\alpha'}~,
\end{equation}
the correct effective hamiltonian for the spin-polarized Fermi gases finally
becomes
\begin{eqnarray}
H^{eff} = \sum_{\alpha} \sum_{\vec{k}}~
       (\epsilon_{\vec{k},\alpha} - \mu'_{\alpha})
         a^{\dagger}_{\vec{k},\alpha} a_{\vec{k},\alpha}
  - V T^{2B}(\vec{0},\vec{0};0) n_{\uparrow} n_{\downarrow}
                                     \hspace*{2.0in} \nonumber \\
  + \frac{1}{V} \sum_{\vec{k} \vec{k}' \vec{q}}~
      T^{MB}_{\uparrow,\downarrow}
        (\vec{0},\vec{0},\vec{k}+\vec{k}';
         \hbar^2(\vec{k}-\vec{k}')^2/4m)~
      a^{\dagger}_{\vec{k}+\vec{q},\uparrow}
      a^{\dagger}_{\vec{k}'-\vec{q},\downarrow}
      a_{\vec{k}',\downarrow} a_{\vec{k},\uparrow}~,
\end{eqnarray}
where we have subtracted the constant
$V T^{2B}(\vec{0},\vec{0};0) n_{\uparrow} n_{\downarrow}$ to avoid a double
counting of the effects of the interaction that have already been included in
the redefinition of the chemical potentials. Notice that we have implicitly
assumed that the substitution $\mu_{\alpha} \rightarrow \mu'_{\alpha}$ is
carried out also in the occupation numbers of Eq.~(\ref{xi}), since this makes
the calculation of the many-body T matrix selfconsistent.

{}From this effective hamiltonian we conclude that the spin-polarized gases are
so-called Fermi liquids \cite{tony}. In particular, the elementary excitations
(or quasiparticles) of the gas have a dispersion relation given by
$\epsilon_{\vec{k},\alpha} - \mu'_{\alpha}$ and scatter of each other with an
amplitude given by the many-body T matrix. As a result the density of atoms in
the two spin states is determined by
\begin{equation}
n_{\alpha} = \frac{1}{V} \sum_{\vec{k}}~N_{\vec{k},\alpha}~,
\end{equation}
with the occupation numbers $N_{\vec{k},\alpha}$ equal to the Fermi
distribution function $(e^{\beta x} + 1)^{-1}$ evaluated at
$\epsilon_{\vec{k},\alpha} - \mu'_{\alpha}$. Moreover, the pressure of the gas
is simply
\begin{equation}
p = \frac{k_BT}{V} \sum_{\alpha} \sum_{\vec{k}}~
     \ln \left( 1 + e^{-\beta(\epsilon_{\vec{k},\alpha}
                              - \mu'_{\alpha})} \right)
     + \frac{4\pi a \hbar^2}{m} n_{\uparrow}n_{\downarrow}
  \equiv \sum_{\alpha} p_{\alpha} + p_{int}~.
\end{equation}

For a positive scattering length $a$, the interaction between the atoms is
effectively repulsive and the second term in the right-hand side is positive.
Interestingly enough, this does not imply that the gas is always mechanically
stable. Indeed, using the hydrodynamic equations
$\partial n_{\alpha}/\partial t
                        = - \nabla \cdot \vec{J}_{\alpha}$ and
$\sum_{\alpha} \partial \vec{J}_{\alpha}/\partial t
                                             = - \nabla p/m$, we find that that
the two sound modes in the gas are described by the coupled wave equations
\begin{equation}
\label{waves}
\frac{\partial^2}{\partial t^2}
 \left[ \matrix{ n_{\uparrow} \cr
                 n_{\downarrow} } \right] = \frac{1}{m}
 \left[ \matrix{ \partial p_{\uparrow}/n_{\uparrow} &
                     \partial p_{int}/n_{\downarrow} \cr
                 \partial p_{int}/n_{\uparrow} &
                     \partial p_{\downarrow}/n_{\downarrow} }
      \right] \cdot
\nabla^2 \left[ \matrix{ n_{\uparrow} \cr
                         n_{\downarrow} } \right]~.
\end{equation}
We thus conclude that the velocities of the sound modes are equal to the square
root of the eigenvalues of the matrix in the right-hand-side of
Eq.~(\ref{waves}), and that both these eigenvalues need to be positive for the
sound velocities to be real and the gas to be mechanically stable.

In the case of a nondegenerate Fermi gas we can apply Boltzmann statistics to
obtain $p_{\alpha} = n_{\alpha} k_BT$. The condition of mechanical stability
then reduces to
$n_{\uparrow} n_{\downarrow} a^6 \le (a/\Lambda)^4/4$ and requires rather low
densities. In the more interesting case of a degenerate Fermi gas, however, we
have
\begin{equation}
p_{\alpha} = n_{\alpha} k_BT \frac{1}{5}
      \left( 3 \sqrt{\frac{\pi}{2}} n_{\alpha}\Lambda^3
                                         \right)^{2/3}~,
\end{equation}
neglecting terms of order $O(1/(k_{F,\alpha}\Lambda)^4)$. The condition on the
densities then becomes
$n_{\uparrow} n_{\downarrow} a^6 \le \pi^2/2304$, which is much less
restrictive. Note that in the mechanically unstable region, the gas is in
general unstable against density fluctuations and will phase separate into a
dense (liquid or solid) and a dilute (gas) phase. This process of phase
separation is known as a spinodal decomposition and the density and temperature
conditions at which it first occurs, lie in our case on a so-called spinodal
surface.

Apart from the first-order phase transition implied by the above instability,
we do not expect any other phase transition to occur in the gas on the basis of
our effective hamiltonian. In particular, we do not expect a phase transition
due to quantum degeneracy, because for that we need an attractive interaction
between the quasiparticles as will be explained shortly. In principle, however,
it is possible that a positive s-wave scattering amplitude induces a negative
p-wave scattering amplitude due to the Kohn-Luttinger effect \cite{kohn}. In
this manner a phase transition to a superfluid state might occur after all.
Although it is interesting to analyze this possibility in more detail, we will
not do so here since it is beyond the scope of the present paper.

\subsection{The case ${\bf a<0}$}
\label{Fl0}
The prospects for achieving a phase transition in the degenerate regime are
more promising for spin-polarized Fermi gases with a negative scattering
length, which actually applies to both atomic deuterium \cite{vianney} and
atomic $^6$Li \cite{marianne}. Nevertheless, for this phase transition to be
experimentally observable, it must take place in the (meta)stable region of the
phase diagram where the gas is mechanically stable. Since the stability
analysis of the previous section is insensitive to the sign of scattering
length, this implies in the degenerate regime that the densities of the two
spin states have to fulfill the requirement $n_{\uparrow} n_{\downarrow} a^6
\le \pi^2/2304$. In contrast to the case of a Fermi gas with a positive
scattering length, the latter condition now does not rule out the possibility
of the formation of a superfluid state at sufficiently low temperatures.

The physical reason for this is that with effectively attractive interatomic
interactions the gas can form a Bose condensate of Cooper pairs, in the same
way as an electron gas can form a condensate of Cooper pairs in the BCS theory
of superconductivity \cite{tony}. Mathematically, the instability of the gas
towards the creation of Cooper pairs is signalled by the fact that the
scattering amplitude diverges for two quasiparticles at the Fermi energy
$\epsilon_F \equiv \hbar^2 k_F^2/2m =
 (\hbar^2 k_{F,\uparrow}^2/2m + \hbar^2 k_{F,\downarrow}^2/2m)/2$ with opposite
momenta and `spin' \cite{thouless}. Using Eqs.~(\ref{Txi}) and (\ref{xi}) and
neglecting the small imaginary part of $\Xi(\vec{0};2\epsilon_F)$, we thus find
that the critical temperature of the BCS transition obeys
\begin{equation}
\label{gap}
\frac{1}{T^{MB}_{\uparrow,\downarrow}
                          (\vec{0},\vec{0},\vec{0};2\epsilon_F)}
  = \frac{m}{4\pi a\hbar^2}
    + {\cal P} \int \frac{d\vec{k}}{(2\pi)^3}~
        \frac{N_{\uparrow}(\vec{k}) + N_{\downarrow}(\vec{k})}
             {2\epsilon_F - \hbar^2\vec{k}^2/m}
  = 0~,
\end{equation}
where ${\cal P}$ denotes the Cauchy principal-value part of the integral.

The properties of this so-called linearized BCS gap equation are well-known
from the theory of superconductivity. First, this equation only has a solution
if $a<0$. Second, the highest critical temperature for a given total density
$n$ is obtained if $n_{\uparrow} = n_{\downarrow}$ or equivalently if
$\delta\epsilon_F = |\hbar^2 k_{F,\uparrow}^2/2m
                          - \hbar^2 k_{F,\downarrow}^2/2m| = 0$.
This can be understood most easily by noting that for a nonzero value of
$\delta\epsilon_F$ there are less particles at the Fermi surface that can be
paired, which suppresses the critical temperature. Third, if the densities in
both spin states are equal, the gap equation can be solved analytically
\cite{mohit} and leads to
\begin{equation}
\label{tc}
T_c \simeq \frac{5\epsilon_F}{3k_B}
              \exp \left\{ -\frac{\pi}{2k_F|a|} - 1 \right\}~.
\end{equation}
For unequal densities an analytical treatment is not feasible, but a numerical
solution of Eq.~(\ref{gap}) shows that nonzero critical temperatures are only
possible if the `spin polarization'
$|n_{\uparrow} - n_{\downarrow}|/(n_{\uparrow} + n_{\downarrow})$
is less than $3k_BT_c/2\epsilon_F$ \cite{marianne}. Qualitatively, this is a
result of the fact that if $\delta\epsilon_F$ is of the order of $k_BT_c$, it
is no longer energetically favorable to form Cooper pairs and the gas is in a
normal state even at zero temperature.

{}From an experimental point of view, we thus conclude that the most favorable
conditions for the observation of the BCS transition in a spin-polarized Fermi
gas are obtained if $n_{\uparrow} = n_{\downarrow} = n/2$. Under these
conditions the mechanical stability of the gas at the temperatures of interest
requires that $n|a|^3 \le \pi/24$ or equivalently that
$k_F|a| \le \pi/2$. For atomic deuterium the triplet scattering length is equal
to $a \simeq -6.8~a_0$ ($a_0$ is the Bohr radius). At realistic densities of
say $10^{12}~cm^{-3}$ we thus find that $k_F|a| \simeq 1.1 \times 10^{-3} \ll
\pi/2$ and the stability condition is easily fulfilled. However, from the
exponential dependence in Eq.~(\ref{tc}) we immediately see that the critical
temperature is completely out of reach with present-day cooling techniques.
Fortunately, this conclusion does not hold for atomic $^6$Li because the
triplet scattering length now has the anomalously large value of
$a \simeq -4.6 \times 10^3~a_0$. For the same density of atoms we  thus find
that $k_F|a| \simeq 0.74 < \pi/2$ and that the critical temperature is only
$29~nK$, which is rather close to the temperatures that recently have been
obtained with the bosonic isotope $^7$Li \cite{rice}. Therefore, it appears
that $^6$Li is a very promising candidate for the observation of the BCS
transition in a weakly interacting gas.

Because of this possible application, we also want to mention briefly how we
can arrive at an accurate description of the gas below the critical
temperature. Although it is not difficult to treat the general situation with
$n_{\uparrow} \neq n_{\downarrow}$, we concentrate here on the most important
case of an equal population of the two hyperfine states. Moreover, we also only
discuss the derivation of the elementary excitations of the superfluid state
(which is sufficient for the most interesting equilibrium properties of the
gas) and do not consider the effective interaction between these
quasiparticles. Introducing the convenient notation
$\mu' \equiv \mu'_{\alpha} - \epsilon_{\alpha}$ for the renormalized chemical
potential (or Fermi energy $\epsilon_F$) of both spin states, our task is
therefore reduced to including into the free hamiltonian of the quasiparticles,
i.e.\ into
\begin{equation}
H_0^{eff} \equiv \sum_{\alpha} \sum_{\vec{k}}~
       (\epsilon_{\vec{k}} - \mu')
         a^{\dagger}_{\vec{k},\alpha} a_{\vec{k},\alpha}
  - V T^{2B}(\vec{0},\vec{0};0) \frac{n^2}{4}~,
\end{equation}
the effect of the Cooper pairs.

To achieve this, we must realize that BCS theory is roughly speaking the theory
of a Bose-Einstein condensation of Cooper pairs. Therefore, the order parameter
of the phase transition is the expectation value of the annihilation operator
for a pair of atoms. (See Sec.~\ref{bose} for a treatment of Bose-Einstein
condensation in an atomic Bose gas.) As a result, we do not only have
nonvanishing average occupation numbers
$\langle a^{\dagger}_{\vec{k},\alpha} a_{\vec{k},\alpha} \rangle$
but also nonvanishing values for
$\langle a_{\vec{k},\alpha} a_{-\vec{k},\alpha'} \rangle$. Note that in our
case we are dealing with a pairing between atoms with different spin states and
the latter average is only nonzero for
$\alpha \neq \alpha'$. To see what the influence of the anomalous averages
$\langle a_{\vec{k},\uparrow} a_{-\vec{k},\downarrow} \rangle$
is, we briefly return to the hamiltonian of Eq.~(\ref{hamilton}). Taking for
the sake of clarity $V_{\vec{q}} = V_{\vec{0}}$, the lowest order average
correction on the hamiltonian $H_0$ due to the interaction now becomes (cf.\
Eq.~(\ref{shift}))
\begin{equation}
\label{change}
\Delta H_0 = \sum_{\alpha} \sum_{\vec{k}}~
    a^{\dagger}_{\vec{k},\alpha} a_{\vec{k},\alpha}
      \left\{ \sum_{\alpha' \neq \alpha}~
        V_{\vec{0}}~ n_{\alpha'} \right\}
    + \sum_{\vec{k}}~
      \left\{ \Delta_0 a^{\dagger}_{\vec{k},\uparrow}
                       a^{\dagger}_{-\vec{k},\downarrow}
            + \Delta^*_0 a_{-\vec{k},\downarrow}
                         a_{\vec{k},\uparrow} \right\}~,
\end{equation}
if we define the BCS gap parameter $\Delta_0$ by
\begin{equation}
\label{orderp}
\Delta_0 \equiv \frac{1}{V} \sum_{\vec{k}}~
             V_{\vec{0}} \langle a_{-\vec{k},\downarrow}
                                 a_{\vec{k},\uparrow}  \rangle~.
\end{equation}
{}From our previous discussion of the normal state of the gas we know that we
can to a good approximation account for all two-body processes by replacing in
Eq.~(\ref{change}) the potential $V_0$ by $T^{2B}(\vec{0},\vec{0};0)$. It is
important, however, that we do not perform this substitution in
Eq.~(\ref{orderp}) because the BCS theory is already going to sum all the other
relevant two-body processes for us \cite{kleinert}.

We thus arrive at the conclusion that the free hamiltonian of the
quasiparticles, including the effects of a condensate of Cooper pairs, is given
by
\begin{eqnarray}
H_0^{eff} = \sum_{\alpha} \sum_{\vec{k}}~
       (\epsilon_{\vec{k}} - \mu')
         a^{\dagger}_{\vec{k},\alpha} a_{\vec{k},\alpha}
  + \sum_{\vec{k}}~
      \left\{ \Delta_0 a^{\dagger}_{\vec{k},\uparrow}
                       a^{\dagger}_{-\vec{k},\downarrow}
            + \Delta^*_0 a_{-\vec{k},\downarrow}
                         a_{\vec{k},\uparrow} \right\}
                                  \hspace*{0.5in} \nonumber \\
  - V \frac{|\Delta_0|^2}{V_{\vec{0}}}
  - V T^{2B}(\vec{0},\vec{0};0) \frac{n^2}{4}~,
\end{eqnarray}
where we have again subtracted a constant term to avoid a double counting of
the interaction effects. Being quadratic in the creation and annihilation
operators, this hamiltonian can be diagonalized by means of a Bogoliubov
transformation and we find that the Bogoliubov quasiparticles have an energy
dispersion
$\hbar \omega_{\vec{k}} =
   \sqrt{(\epsilon_{\vec{k}} - \mu')^2 + |\Delta_0|^2}$
which for $n_{\uparrow} = n_{\downarrow}$ is independent of their `spin'.
Moreover, after this diagonalisation the average density $n$ and order
parameter $\Delta_0$ are easily calculated and result in the equation of state
\begin{equation}
n = \frac{1}{V} \sum_{\alpha} \sum_{\vec{k}}~
    \left\{ \frac{\epsilon_{\vec{k}} - \mu'}
                 {\hbar\omega_{\vec{k}}} N_{\vec{k},\alpha}
      + \frac{\hbar\omega_{\vec{k}} - \epsilon_{\vec{k}} + \mu'}
             {2\hbar\omega_{\vec{k}}} \right\}
\end{equation}
and the BCS gap equation
\begin{equation}
\frac{1}{V_{\vec{0}}} + \frac{1}{V} \sum_{\vec{k}}~
   \frac{1 - N_{\vec{k},\uparrow} - N_{\vec{k},\downarrow}}
        {2\hbar\omega_{\vec{k}}} =0~,
\end{equation}
respectively. Here, the average occupation numbers of the quasiparticles are
again denoted by $N_{\vec{k},\alpha}$, but they are now equal to the Fermi
distribution function evaluated at $\hbar\omega_{\vec{k}}$ and in fact
independent of $\alpha$.

The equation of state and the BCS gap equation are, at a given density and
temperature, two equations for the two unknown quantities $\mu'$ and
$|\Delta_0|$. However, before we can actually solve these equations we first
have to resolve the ultraviolet divergence in the gap equation. This divergence
is a consequence of the fact that we have neglected the momentum dependence of
the potential and used $V_{\vec{q}} = V_{\vec{0}}$. Making use of the same
approximation in the Lippman-Schwinger equation Eq.~(\ref{t2b}) and of the fact
that the Fermi energy $\mu'=\hbar^2k_F^2/2m$ is always much smaller than
$\hbar^2/mr_V^2$, we find that the two-body T matrix obeys
\begin{equation}
\frac{1}{T^{2B}(\vec{0},\vec{0};0)} \simeq
\frac{1}{T^{2B}(\vec{0},\vec{0};2\mu')} =
  \frac{1}{V_{\vec{0}}} + \frac{1}{V} \sum_{\vec{k}}~
     \frac{1}{2(\epsilon_{\vec{k}}-\mu')}
\end{equation}
in this case. We thus observe that the divergence in the BCS gap equation can
be cancelled by a renormalization of $1/V_{\vec{0}}$ to
$1/T^{2B}(\vec{0},\vec{0};0) = m/4\pi a \hbar^2$. In this manner we then obtain
the following gap equation
\begin{equation}
\frac{m}{4\pi a \hbar^2} +
  \frac{1}{V} \sum_{\vec{k}}~
  \left\{
    \frac{1 - N_{\vec{k},\uparrow} - N_{\vec{k},\downarrow}}
         {2\hbar\omega_{\vec{k}}}
    - \frac{1}{2(\epsilon_{\vec{k}}-\mu')}
  \right\} = 0~,
\end{equation}
which is free of divergences. Furthermore, taking the limit $\Delta_0
\rightarrow 0$ to obtain an equation for the critical temperature $T_c$, we
exactly reproduce (after taking also the thermodynamic limit and substituting
$\mu'=\epsilon_F$) the result of Eq.~(\ref{gap}) which shows the consistency of
our approach and, in particular, also that the BCS theory in essence indeed
sums all two-body scattering processes.

\section{BOSE GASES}
\label{bose}
We now turn our attention to the atomic Bose gases, which have caused so much
excitement in the atomic physics community after the experimental observation
of Bose-Einstein condensation in atomic $^{87}$Rb, $^7$Li and $^{23}$Na
vapours. As in these experiments, we consider here only the doubly
spin-polarized case. This implies that there is only one spin state in the
problem and we will, therefore, in this section completely suppress the spin
indices in our notation.

\subsection{The case ${\bf a>0}$}
\label{Bg0}
Again, we first consider a gas with an effectively repulsive triplet
interaction, which applies for example to atomic hydrogen \cite{H}, atomic
$^{87}$Rb \cite{rb} and atomic $^{23}$Na \cite{na}. The appropriate effective
hamiltonian for the doubly spin-polarized atomic Bose gases is, in analogy with
the discussion in Sec.~\ref{Fg0}, given by
\begin{eqnarray}
\label{heff}
H^{eff} = \sum_{\vec{k}}~
       (\epsilon_{\vec{k}} - \mu') a^{\dagger}_{\vec{k}}
                                                  a_{\vec{k}}
  - V T^{2B}(\vec{0},\vec{0};0) n^2 \hspace*{2.0in} \nonumber \\
  + \frac{1}{2V} \sum_{\vec{k} \vec{k}' \vec{q}}~
      T^{MB}(\vec{0},\vec{0},\vec{k}+\vec{k}';
             \hbar^2(\vec{k}-\vec{k}')^2/4m)~
      a^{\dagger}_{\vec{k}+\vec{q}}
      a^{\dagger}_{\vec{k}'-\vec{q}}
      a_{\vec{k}'} a_{\vec{k}}~,
\end{eqnarray}
with a renormalized chemical potential $\mu'$ equal to
\begin{equation}
\label{mu}
\mu' = \mu - 2nT^{2B}(\vec{0},\vec{0};0)~.
\end{equation}
Looking at this hamiltonian we immediately see that the density of atoms is
determined by
\begin{equation}
\label{density}
n = \frac{1}{V} \sum_{\vec{k}} N_{\vec{k}}~
\end{equation}
where the average occupation numbers $N_{\vec{k}}$ are now equal to the Bose
distribution function $(e^{\beta x} - 1)^{-1}$ evaluated at $\epsilon_{\vec{k}}
- \mu'$. In addition, the pressure of the gas is
\begin{equation}
\label{press}
p = - \frac{k_BT}{V} \sum_{\vec{k}}~
     \ln \left( 1 - e^{-\beta(\epsilon_{\vec{k}}
                              - \mu')} \right)
     + \frac{4\pi a \hbar^2}{m} n^2~,
\end{equation}
which for a gas with a positive scattering length does not show a mechanical
instability, i.e.\ $dp/dn$ is always greater than zero for densities such that
$nr_V^3 \ll 1$ and our theory is applicable.

On the basis of Eq.~(\ref{density}) we thus expect the gas to Bose condense at
the same critical temperature
\begin{equation}
\label{t0}
T_c = T_0 \equiv \frac{2\pi\hbar^2}{mk_B}
                   \left( \frac{n}{\zeta(3/2)} \right)^{2/3}
\end{equation}
as the ideal Bose gas. For an interacting Bose gas with positive scattering
length the order parameter corresponding to this phase transition is the
expectation value of the annihilation operator for an atom \cite{huang}. If we
want to describe the gas also below the critical temperature, we therefore need
to consider the influence of a nonvanishing average $\langle a_{\vec{0}}
\rangle$ on the effective hamiltonian. Focussing again on the derivation of the
elementary excitations in the superfluid state, this is most easily achieved by
first carrying out the substitution
$a_{\vec{k}} \rightarrow
 \delta_{\vec{k},\vec{0}} \langle a_{\vec{0}} \rangle +
                                                   a_{\vec{k}}$
in the hamiltonian $H^{eff}$ and then keeping only all terms that are of first
or second order in the creation and annihilation operators. In this manner we
find in a good approximation that
\begin{eqnarray}
\label{hquasi}
H_0^{eff} &=&
 - V \left(
        T^{2B}(\vec{0},\vec{0};0) (n')^2
        + T^{MB}(\vec{0},\vec{0},\vec{0};0) n' n_0
        + \frac{1}{2} T^{MB}(\vec{0},\vec{0},\vec{0};0) n_0^2
     \right)                                      \nonumber \\
 &+& \sqrt{Vn_0}
     \left( -\mu' + n_0 T^{MB}(\vec{0},\vec{0},\vec{0};0)
     \right)
       \left( a_{\vec{0}} + a^{\dagger}_{\vec{0}} \right)
  + \sum_{\vec{k}}~(\epsilon_{\vec{k}} - \mu')
                           a^{\dagger}_{\vec{k}} a_{\vec{k}}
                                                 \nonumber \\
 &+& \sum_{\vec{k}}~\left\{
     \left( 2n_0T^{MB}(\vec{0},\vec{0},\vec{0};0) \right)
        a^{\dagger}_{\vec{k}} a_{\vec{k}}
     + \frac{1}{2} n_0 T^{MB}(\vec{0},\vec{0},\vec{0};0)
          \left( a^{\dagger}_{\vec{k}} a^{\dagger}_{-\vec{k}}
                 + a_{-\vec{k}} a_{\vec{k}} \right) \right\}~,
\end{eqnarray}
if we choose $\langle a_{\vec{0}} \rangle$ to be real and introduce the
condensate density
$n_0 \equiv \langle a_{\vec{0}} \rangle^2/V$ and the density of noncondensed
atoms $n' = n - n_0$. The most difficult part of the calculation is again to
avoid double countings of the interaction effects. Besides the subtractions in
the first term in the right-hand-side, this also requires that the relation
between $\mu$ and $\mu'$ is changed into
$\mu = 2n'T^{2B}(\vec{0},\vec{0};0) + \mu'$, which reduces to Eq.~(\ref{mu}) if
the condensate density vanishes.

To proceed, we first note that if we want
$\langle a_{\vec{0}} \rangle$ to be the total expectation value of the original
operator $a_{\vec{0}}$, we must require that the expectation value of the new
operator $a_{\vec{0}}$ (i.e.\ after the substitution) disappears. This is
achieved by putting
$\mu' = n_0 T^{MB}(\vec{0},\vec{0},\vec{0};0)$, because then the terms in the
hamiltonian linear in the creation and annihilation operators vanish. As a
result, the chemical potential is determined by
\begin{equation}
\mu = 2(n-n_0) T^{2B}(\vec{0},\vec{0};0) +
            n_0 T^{MB}(\vec{0},\vec{0},\vec{0};0)~,
\end{equation}
which for the approximation that we are using is just the famous
Hugenholtz-Pines relation \cite{hugenholtz}. Because our theory obeys this
relation, we expect that the dispersion relation $\hbar\omega_{\vec{k}}$ of the
quasiparticles is linear at long wavelengths. This indeed turns out to be the
case, since a diagonalization of the above hamiltonian by means of a Bogoliubov
transformation shows that
$\hbar\omega_{\vec{k}} =
\sqrt{ \epsilon_{\vec{k}}^2
  + 2n_0 T^{MB}(\vec{0},\vec{0},\vec{0};0) \epsilon_{\vec{k}} }$
\cite{bogol} and that
\begin{eqnarray}
H_0^{eff} =
  \sum_{\vec{k}}~\hbar\omega_{\vec{k}}
                           b^{\dagger}_{\vec{k}} b_{\vec{k}}
                                 \hspace*{4.0in} \nonumber \\
 - V \left(
        T^{2B}(\vec{0},\vec{0};0) (n')^2
        + T^{MB}(\vec{0},\vec{0},\vec{0};0) n' n_0
        + \frac{1}{2} T^{MB}(\vec{0},\vec{0},\vec{0};0) n_0^2
     \right)~,
\end{eqnarray}
where $b^{\dagger}_{\vec{k}}$ and $b_{\vec{k}}$ denote the creation and
annihilation operators for the Bogoliubov quasiparticles and we have neglecting
between the brackets additional terms that are a factor of order
$O(\sqrt{n_0a^3})$ smaller \cite{huang}. Moreover, using this Bogoliubov
transformation we can now obtain also for the superfluid state the pressure
\begin{eqnarray}
p = - \frac{k_BT}{V} \sum_{\vec{k}}~
     \ln \left( 1 - e^{-\beta\hbar\omega_{\vec{k}}} \right)
                                   \hspace*{3.0in} \nonumber \\
     + T^{2B}(\vec{0},\vec{0};0) (n')^2
     + T^{MB}(\vec{0},\vec{0},\vec{0};0) n' n_0
     + \frac{1}{2} T^{MB}(\vec{0},\vec{0},\vec{0};0) n_0^2
\end{eqnarray}
and the equation of state
\begin{equation}
n = n_0 + \sum_{\vec{k}}~\left\{
    \frac{\epsilon_{\vec{k}}
           + n_0 T^{MB}(\vec{0},\vec{0},\vec{0};0)}
         {\hbar\omega_{\vec{k}}} N_{\vec{k}}
    + \frac{\epsilon_{\vec{k}}
            + n_0 T^{MB}(\vec{0},\vec{0},\vec{0};0)
            - \hbar\omega_{\vec{k}}}
           {2\hbar\omega_{\vec{k}}} \right\}~,
\end{equation}
which determines the condensate density at a fixed density and temperature.
Note that in the equation of state the average occupation numbers of the
quasiparticles $N_{\vec{k}}$ are equal to the Bose distribution function
evaluated at $\hbar\omega_{\vec{k}}$.

This almost completes our discussion of the interacting Bose gas with a
positive scattering length. There, however, remains one point that we need to
address, namely our theory is not selfconsistent yet. This is because we have
not yet mentioned how the many-body T matrix is determined below the critical
temperature $T_c$. In general, this is a complicated problem due to the
presence of infrared divergences in the theory of a dilute Bose gas \cite{N}.
At very low temperatures such that $na\Lambda^2 \gg 1$ the treatment of these
divergences requires more advanced methods then the ones we are using here.
Therefore, we consider only the opposite limit $na\Lambda^2 \ll 1$, which is in
fact the most relevant one for experiments. In this regime the average kinetic
energy $k_BT$ of the atoms is much larger than the average interaction energy
$4\pi an\hbar^2/m$ and the Bogoliubov dispersion is for all practical purposes
well approximated by
$\hbar\omega_{\vec{k}} \simeq \epsilon_{\vec{k}}
      + n_0 T^{MB}(\vec{0},\vec{0},\vec{0};0) =
                                    \epsilon_{\vec{k}} + \mu'$. In
Eq.~(\ref{hquasi}) this means that we can neglect the anomalous terms
proportional to
$a^{\dagger}_{\vec{k}} a^{\dagger}_{-\vec{k}}$ or
$a_{-\vec{k}} a_{\vec{k}}$. Hence, we now basically deal with the hamiltonian
$\sum_{\vec{k}}~(\epsilon_{\vec{k}}+\mu')a^{\dagger}_{\vec{k}}
                                                    a_{\vec{k}}$
and the many-body T matrix is again just given by Eqs.~(\ref{Txi}) and
(\ref{xi}). Finally, we have to deal with one last subtlety. In the way we have
defined the many-body T matrix, it describes the scattering of quasiparticles.
However, we are dealing with a Bose condensate of particles and we actually
need their scattering amplitude. In our case this, fortunately, only leads to a
slight shift
$E \rightarrow E - 2\mu'$ in the energy of the many-body T matrix. We thus
arrive at a fully selfconsistent theory of the superfluid phase in the regime
$na\Lambda^2 \ll 1$, if in all formulas below Eq.~(\ref{t0}),
$T^{MB}(\vec{0},\vec{0},\vec{0};0)$ is replaced by the many-body T matrix
$T^{MB}(\vec{0},\vec{0},\vec{0};-2\mu')$.

The transition matrix $T^{MB}(\vec{0},\vec{0},\vec{0};-2\mu')$ is almost equal
to $T^{2B}(\vec{0},\vec{0};0) = 4\pi a\hbar^2/m$ for most temperatures below
$T_c$, but decreases to zero for temperatures near the critical temperature due
to the fact that $\Xi(\vec{0};-2\mu')$ diverges in the limit $\mu' \rightarrow
0$. Therefore, our selfconsistent approach reduces to the conventional
Bogoliubov (or Popov) theory of a weakly interacting Bose gas for most
temperatures. However, near the critical temperature it resolves a
long-standing problem of the Bogoliubov theory, which actually predicts a
first-order phase transition \cite{fo} instead of the second-order phase
transition expected from the theory of critical phenomena. Of course, using the
many-body T matrix instead of the two-body T matrix does not include all the
critical fluctuations that are of importance near the critical temperature.
Nevertheless, using the many-body T matrix we incorporate, at least
qualitatively correctly, the fact (known from renormalization group theory
\cite{zinn}) that the interaction between the condensate particles should
vanish exactly at the critical temperature. It is rather interesting that this
can already be achieved by including only all two-body scattering processes.

\subsection{The case ${\bf a<0}$}
\label{Bl0}
We now turn to the physics of a Bose gas with effectively attractive
interactions, which is realized for instance in atomic $^7$Li \cite{li} and
atomic $^{133}$Cs \cite{cs}. This is an interesting topic because a thoughtless
application of the theory of Sec.~\ref{Bg0} would result in a dispersion of the
Bogoliubov quasiparticles that is equal to
$\hbar\omega_{\vec{k}} = \sqrt{ \epsilon_{\vec{k}}^2
                 - (8\pi |a|n_0\hbar^2/m) \epsilon_{\vec{k}} }$
and therefore purely imaginary at long wavelengths. From this simple argument
we can thus already conclude that Bose-Einstein condensation in the canonical
sence, i.e.\ with the order parameter $\langle a_{\vec{0}} \rangle$, will not
occur if the scattering length is negative \cite{henk}. However, from the
experience with fermionic gases gained in Sec.~\ref{Fl0}, we might expect a
BCS-like phase transition to take place instead.    This would be signalled in
Eq.~(\ref{heff}) by a divergence of the scattering amplitude of two
quasiparticles at energy $\mu'$ and with opposite momenta, or by (cf.\
Eq.~(\ref{gap}))
\begin{equation}
\frac{1}{T^{MB}(\vec{0},\vec{0},\vec{0};2\mu')} =
  \frac{1}{T^{2B}(\vec{0},\vec{0};0)} -
  \int \frac{d\vec{k}}{(2\pi)^3}~
          \frac{N(\vec{k})}{\mu' - \hbar^2\vec{k}^2/2m} = 0~
\end{equation}
with of course $N(\vec{k})$ the Bose distribution function evaluated at
$\hbar^2 \vec{k}^2/2m - \mu'$. Analyzing this condition, we find that the phase
transition indeed occurs at a temperature $T_{BCS} = T_0( 1 +
O(|a|/\Lambda_0))$ which is only slightly above the critical temperature of the
ideal Bose gas.

Following the same procedure as in Sec.~\ref{Fl0}, we can easily show that
below this critical temperature the equation of state becomes
\begin{equation}
\label{n1}
n = \frac{1}{V} \sum_{\vec{k}}~
    \left\{ \frac{\epsilon_{\vec{k}} - \mu'}
                 {\hbar\omega_{\vec{k}}} N_{\vec{k}}
      + \frac{\epsilon_{\vec{k}} - \mu' - \hbar\omega_{\vec{k}}}
             {2\hbar\omega_{\vec{k}}} \right\}
\end{equation}
and that the BCS gap equation is well approximated by
\begin{equation}
\label{g1}
\frac{m}{4\pi a\hbar^2} + \frac{1}{V} \sum_{\vec{k}}~
            \frac{N_{\vec{k}}}{\hbar\omega_{\vec{k}}} = 0~.
\end{equation}
Here the energy of the Bogoliubov quasiparticles is
$\hbar\omega_{\vec{k}} =
        \sqrt{ (\epsilon_{\vec{k}}-\mu')^2 - |\Delta_0|^2 }$
and their average occupation numbers $N_{\vec{k}}$ are equal to the Bose
distribution function evaluated at this energy. These equations again determine
$\mu'$ and $|\Delta_0|$ at a fixed density and temperature.

Notice that in contrast with the fermion case, the dispersion
$\hbar\omega_{\vec{k}}$ has a minus sign in front of $|\Delta_0|^2$. This has
important consequences because it implies that if we lower the temperature at a
fixed density, both    $|\Delta_0|^2$ and $\mu'$ increase in such a manner that
the gap in the Bogoliubov dispersion decreases. Hence, at a second temperature
$T_{BEC}<T_{BCS}$ the gap closes and the number of particles in the
zero-momentum state diverges, which signals a Bose-Einstein condensation. Below
that second critical temperature we always have $|\Delta_0| = -\mu'$ and the
dispersion $\hbar\omega_{\vec{k}}$ becomes equal to
$\sqrt{\epsilon_{\vec{k}}^2 - 2\mu'\epsilon_{\vec{k}}}$ which, interestingly
enough, is exactly the same Bogoliubov dispersion as for a Bose-condensed gas
with positive scattering length. Moreover, Eqs.~(\ref{n1}) and (\ref{g1}) turn
into
\begin{equation}
n = n_0 + \frac{1}{V} \sum_{\vec{k} \neq \vec{0}}~
    \left\{ \frac{\epsilon_{\vec{k}} - \mu'}
                 {\hbar\omega_{\vec{k}}} N_{\vec{k}}
      + \frac{\epsilon_{\vec{k}} - \mu' - \hbar\omega_{\vec{k}}}
             {2\hbar\omega_{\vec{k}}} \right\}
\end{equation}
and
\begin{equation}
\frac{m}{4\pi a\hbar^2} - \frac{n_0}{\mu'}
  + \frac{1}{V} \sum_{\vec{k} \neq \vec{0}}~
            \frac{N_{\vec{k}}}{\hbar\omega_{\vec{k}}} = 0~,
\end{equation}
determining now $\mu'$ and the condensate density $n_0$. It is important to
mention here that the possibility of a Bose-Einstein condensation in a gas with
effectively attractive interactions, in the way that we have just seen, does
not contradict the argument against Bose-Einstein condensation presented in the
beginning of this section. The reason is that in that argument we assumed that
a Bose-Einstein condensation would be associated with the order parameter
$\langle a_{\vec{0}} \rangle$ and did not consider the possibility that it
would be associated with the condensate density $n_0$, as in the ideal Bose gas
\cite{henk}.

It thus appears that a Bose gas with a negative scattering length offers the
exciting possibility of observing both a BCS transition and a Bose-Einstein
condensation. However, we have up to now not considered the mechanical
stability of the gas in these two superfluid phases. This is clearly of the
utmost importance, because Eq.~(\ref{press}) shows that for a negative
scattering length $dp/dn$ does not always have to be positive. Indeed,
calculating the spinodal line from $dp/dn = 0$, we actually find the
unfortunate result that the gas is always unstable at $T_{BCS}$. We are thus
forced to conclude that the two second-order phase transitions that we have
found above are always preempted by a first-order gas-liquid or gas-solid
transition and are, therefore, in a homogeneous system unobservable.

\section{DISCUSSION AND OUTLOOK}
\label{out}
In this paper we have presented a unified picture of weakly interacting atomic
gases, focussing on the differences that arise due to the statistics of the
atoms and due to the interatomic interaction being either effectively repulsive
or attractive. However, we have restricted our discussion to the homogeneous
case, whereas the experiments are up till now always performed in an
inhomogeneous magnetic trap. Fortunately, the experiments are also always
performed at temperatures such that $k_BT \gg \hbar\omega$, where $\hbar\omega$
is the energy splitting of the (harmonic oscillator) trapping potential. Under
these conditions we are generally justified in using a local-density
approximation, which means that we consider the gas as homogeneous in each
point in space. In this sense the theory presented above can also be applied to
an inhomogeneous situation and we, in first instance, do not anticipate any
qualitative differences. Nevertheless, for quantitative predictions of for
example the critical temperature and the density profile in the trap, it is of
course essential to include the effect of the trapping potential.

Moreover, for a Bose gas with negative scattering length there is also a
qualitative difference. In Sec.~\ref{Bl0} we argued that for a homogeneous gas
a Bose condensate is always unstable in this case. However, first Hulet
\cite{hulet} and subsequently Ruprecht {\it et al.} \cite{keith} have argued
that in an inhomogeneous situation one can have a metastable condensate if the
number of particles is sufficiently small. This was recently put on a more firm
theoretical basis by Kagan {\it et al.} \cite{kagan} and by one of us
\cite{stoof}. In particular, the last author has shown by a calculation of the
decay rate of the condensate due to thermal fluctuations, that a long-lived
metastable condensate can exist for temperatures sufficiently close to the
critical temperature. Although this result appears to explain the recent
observation of Bose-Einstein condensation in atomic $^7$Li, our theoretical
understanding of these experiments is still incomplete. The most important
problem that remains to be solved is the stability of the gas just above the
critical temperature. Clearly, on the basis of a local-density approximation we
would conclude, as in Sec.~\ref{Bl0}, that the gas is unstable and that also in
an inhomogeneous situation Bose-Einstein condensation is unobservable. However,
the local-density approximation is invalid near the critical temperature due to
the large value of the (homogeneous) correlation length. A thorough discussion
of this point, therefore, requires a theory that goes beyond this approximation
and is evidently of great interest in view of the Rice experiments with $^7$Li
\cite{rice}.

Also in the case of a Bose gas with positive scattering length, interesting
theoretical problems remain. One of them is the correct treatment of the
infrared divergences in the regime $na\Lambda^2 \gg 1$, which, due to the
effectiveness of evaporative cooling techniques, is also anticipated to be
experimentally accessible. As already mentioned, we presumably need more
advanced methods than discussed here to accurately perform a resummation of the
divergent contributions to the thermodynamic quantities of interest. Indeed, we
are presently carrying out a renormalization group study of the interacting
Bose gas in this regime, because by this method infrared divergences are
automatically avoided. Moreover, a resolution of these divergences is also of
importance for the theory of two-dimensional atomic Bose gases, such as for
instance doubly spin-polarized atomic hydrogen adsorbed on superfluid $^4$He
films. The reason for this is that in a two-dimensional geometry, the
divergences are so important that they actually prevent the possibility of a
Bose condensate (in the thermodynamic limit). Nevertheless, the gas is expected
to undergo a phase transition to a superfluid state at sufficiently low
temperatures. In our opinion, formulating an accurate microscopic theory
\cite{michel} for this so-called Kosterlitz-Thouless transition is an important
challenge for the future.

Finally, we would also like to mention the possibility of studying mixtures of
fermionic and bosonic atomic gases. One exciting possibility is a mixture of
doubly spin-polarized atomic $^6$Li and $^7$Li. In this manner one would create
the dilute analogue of $^3$He and $^4$He mixtures. Such a mixture of doubly
spin-polarized atomic $^6$Li and $^7$Li is rather interesting because, as we
have seen in Sec.~\ref{fermi}, the $^6$Li atoms essentially do not interact
with each other but only with the $^7$Li atoms. However, due to the latter
interaction they can now also be evaporatively cooled on a timescale which is
much smaller than their lifetime. It therefore seems worthwhile to extend the
methods discussed in this paper to this system and in particular study
Bose-Einstein condensation in the presence of an essentially ideal Fermi gas.

It is clear from these examples, and from many others that we unfortunately are
not able to discuss here, that we are heading towards an exciting future, in
which no doubt many surprises will await us. We hope that this paper will be of
some use to people that intend to contribute to this rapidly developing area of
atomic physics.

\section*{ACKNOWLEDGMENTS}
We are very grateful for many illuminating discussions over the years with, in
particular, Keith Burnett, Steve Girvin, Tom Hijmans, Randy Hulet, Tony
Leggett, Meritt Reynolds, Ike Silvera, and Jook Walraven.

\end{document}